\def\year{2020}\relax
\documentclass[letterpaper]{article} 
\usepackage{aaai20}  
\usepackage{times}  
\usepackage{helvet} 
\usepackage{courier}  
\usepackage[hyphens]{url}  
\usepackage{graphicx} 
\usepackage{listings}
\usepackage{graphicx}  
\usepackage{xcolor}
\usepackage{hyperref}
\usepackage{cleveref}
\usepackage{xspace}
\usepackage{amssymb}        
\definecolor{light-gray}{gray}{0.95}
\definecolor{mid-gray}{gray}{0.85}
\definecolor{darkred}{rgb}{0.7,0.25,0.25}
\definecolor{darkgreen}{rgb}{0.15,0.55,0.15}
\definecolor{darkblue}{rgb}{0.1,0.1,0.5}
\definecolor{blue}{rgb}{0.19,0.58,1}

\newcommand{\blue}[1]{\textcolor{blue}{#1}}

\newcommand{\eat}[1]{}

\newcommand{\stitle}[1]{\vspace{2pt}\noindent\textbf{#1}}

\newcommand{\difftree}[0]{\texttt{difftree}\xspace}

\title{ Monte Carlo Tree Search for Generating Interactive Data Analysis Interfaces}

\author{Yiru Chen, Eugene Wu\\
Columbia University\\
yiru.chen@columbia.edu, ewu@cs.columbia.edu }

\begin{document}
\maketitle
\begin{abstract}
Interactive tools like user interfaces help democratize data access for end-users by hiding underlying programming details and exposing the necessary widget interface to users. Since customized interfaces are costly to build, automated interface generation is desirable.  SQL is the dominant way to analyze data and there already exists logs to analyze data. Previous work proposed a syntactic approach to analyze structural changes in SQL query logs and automatically generates a set of widgets to express the changes. However, they do not consider layout usability and the sequential order of queries in the log. We propose to adopt Monte Carlo Tree Search(MCTS) to search for the optimal interface that accounts for hierarchical layout as well as the usability in terms of how easy to express the query log. 
\end{abstract}

\section{Introduction}

SQL is the dominant language for accessing and analyzing large datasets today.  Its expressive power is useful to identify the appropriate queries during ad-hoc analysis (e.g., in a Jupyter notebook).  However, it is cumbersome to repeatedly use for the same set of analysis tasks, and inaccessible to many end-users.  In contrast, customized interactive interfaces help users quickly accomplish their data analysis tasks by hiding underlying programming complexity and exposing a simple set of visual widgets designed for the tasks.  Unfortunately, turning those analysis queries into a reusable interactive interface requires considerable design and programming expertise.  

Prior work~\cite{Zhang2017MiningPI} proposed an automatic interface generation method.  Given a set of analysis queries, it identifies changes between the abstract syntax trees (AST) of the queries (\Cref{f:asts}), and chooses a set of customized interactive widgets (e.g., slider, tabs, buttons) from a predefined library that can express those changes.  For instance, if the queries differ by a numeric value (e.g., \texttt{a=1}, \texttt{a=2}), then it maps the changes (e.g., $1\to 2$) to a widget template (e.g., a slider) that can express the different values. It uses a bottom-up approach that enumerates subtree differences between every pair of ASTs, and maps differences at the same path in the AST to a widget.  

Although this work has shown promise, it still suffers from a number of limitations.  First, it groups subtrees at the same location in the ASTs and matches them to a widget without consideration of the other widgets nor whether the subtrees should be grouped together. Second, it returns a set of widgets that does not account for the interface layout nor constraints such as the screen size.  Notably, it does not leverage the body of HCI research that has studied and quantified interface layout and usability~\cite{Comber1997LayoutCD,Gajos2004SUPPLEAG}.  Third, it ignores the effort needed to use the interface to express the {\it sequence} of input queries.  

To this end, we describe our preliminary work on a top-down search-based approach towards interface generation that explicitly addresses the above limitations.  We propose a \difftree representation of the input query ASTs whose structure also encodes the interface layout---this represents a state in the search space---and define transition rules that incrementally transform the \difftree.  The search space is extremely large, thus we use Monte Carlo Tree Search (MCTS) to efficiently identify the lowest cost interface.  The rest of this paper describes the problem and our current approach, preliminary results, and ongoing research directions.

\section{Problem Overview}

Our goal is to take as input a sequence of SQL queries that are part of an analysis task (e.g., from a query log, or provided by a developer during or after an analysis session), and output an interactive data analysis interface that can express the input queries (and likely similar queries not explicitly in the log).     Our assumption is that the structural differences between the queries are representative of the types of changes the user wishes to express interactively. 

Our approach is to 1) extract syntactic differences between queries, 2) choose interactive widgets that can express those differences as transformations, and 3) design and layout an interative interface.   In this work, we leverage existing automatic visualization techniques that recommend visualizations based on a dataset~\cite{Mackinlay2007ShowMA,sievert2017plotly}, and thus we focus on the joint problem of determining a good layout, and selecting and configuring the appropriate widgets for the layout.   

\begin{figure}
\centering
\includegraphics[width=.9\linewidth]{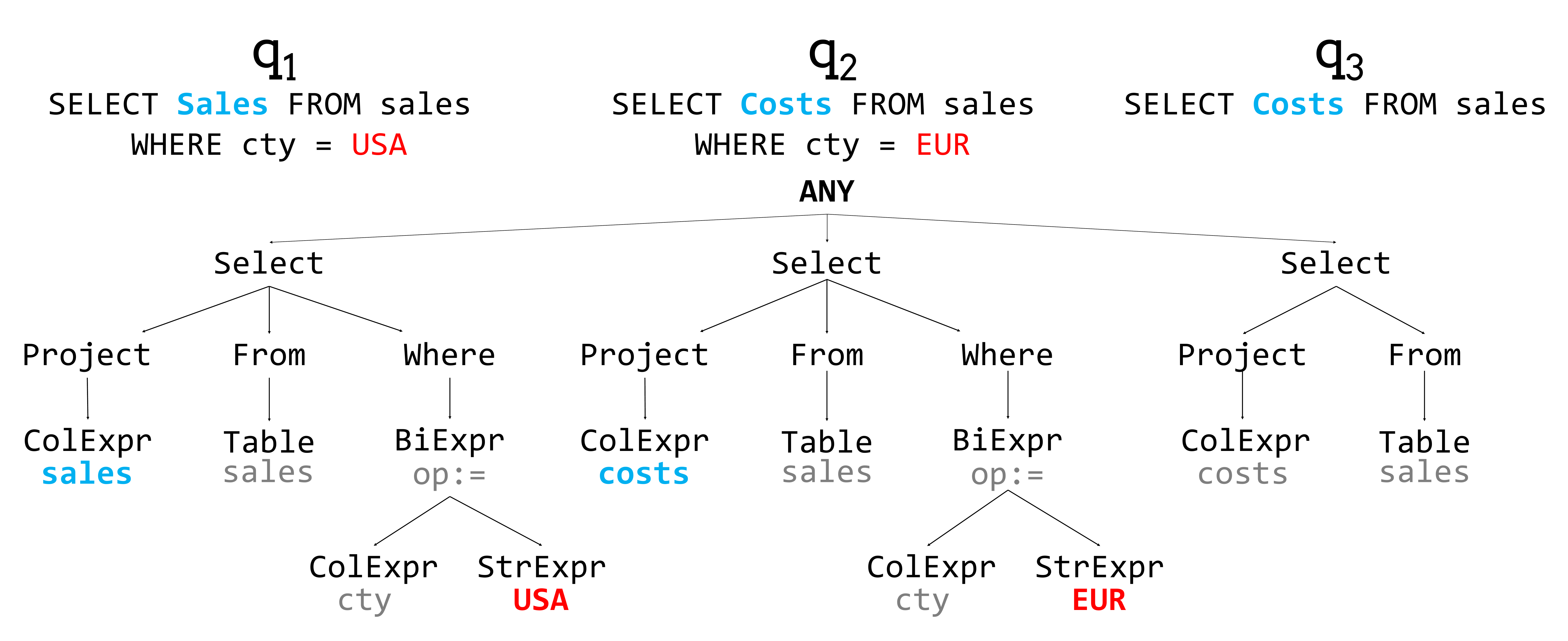}
\caption{\small Example ASTs for 3 SQL queries.}
\label{f:asts}
\vspace{-0.1in}
\end{figure}

\stitle{Queries:} Similar to \cite{Zhang2017MiningPI}, we model each query as its abstract syntax tree (AST).  \Cref{f:asts} without the \texttt{ANY} node illustrates the simplified ASTs of three queries. Each node (e.g., \texttt{Select}, \texttt{BiExpr}) corresponds to a rule in the query grammar.  $q_1$ and $q_2$ differ at two nodes: \texttt{ColExpr} changed from \texttt{sales} to \texttt{costs}, while \texttt{StrExpr} changed from \texttt{USA} to \texttt{EUR}.  $q_3$ differs from $q_2$ by dropping the \texttt{WHERE} clause completely.

\stitle{Interfaces:} An interface is a set of visualizations, a set of interactive widgets, and a hierarchical layout of the visualizations and widgets.  For ease of discussion, we will describe the case where there is a single visualization.    A visualization renders the output of the current query $q$, and each widget changes $q$ based on the user's interactions (e.g., changing a slider, typing in a text box).  When the current query changes, it is re-executed and the results update the visualization.  

\begin{figure}
\centering
\includegraphics[width=\columnwidth]{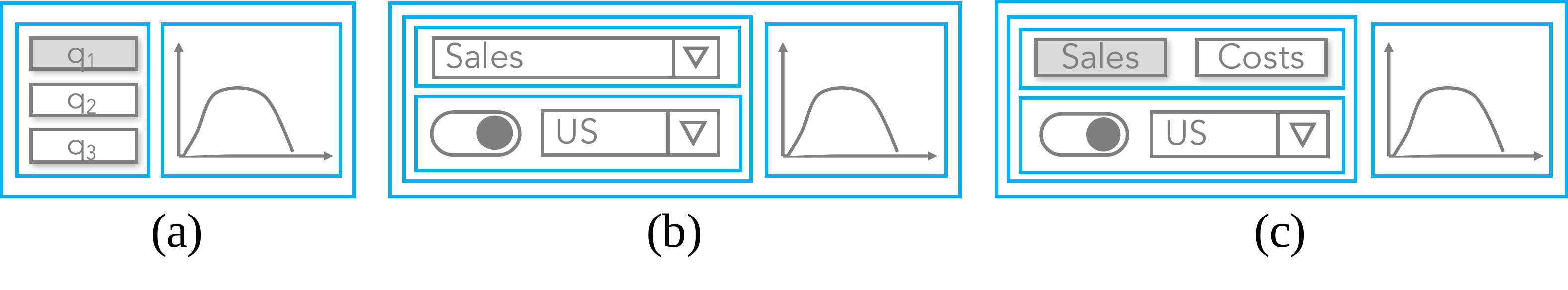}
\vspace{-.2in}
\caption{\small Examples of three interfaces that express the queries in \Cref{f:asts}.  \blue{Blue boxes} depict bounding boxes in the layout hierarchy.}
\label{f:layouts}
\end{figure}

\stitle{Layouts:} \Cref{f:layouts} shows three possible interfaces that can express the queries in \Cref{f:asts}.  The \blue{blue boxes} represent the bounding boxes in the layout hierarchy; for simplicity, we only depict layouts with widgets to the left of the visualization.    For example, \Cref{f:layouts}(a) vertically organizes three buttons, where clicking on a button loads the corresponding query.  (b) uses two dropdowns to change the column and string expressions, respectively, and uses a toggle widget to specify whether the \texttt{WHERE} clause should be in the query.  The toggle and dropdown for the string expression are organized together because they relate to the same parts of the AST.   (c) uses the same layout as (b), but uses the available width to list both column expressions (\texttt{Sales}, \texttt{Costs}) as buttons organized horizontally. 

\begin{figure}[h]
\centering
\includegraphics[width=\columnwidth]{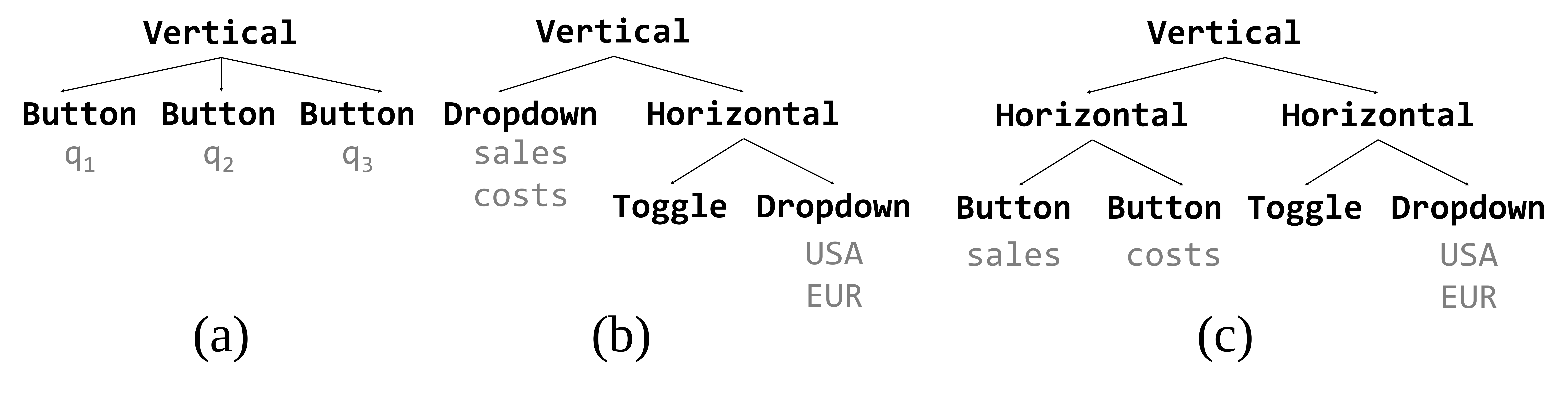}
\vspace{-.3in}
\caption{\small Widget Trees for the interfaces in \Cref{f:layouts}.} 
\label{f:uitree}
\vspace{-.1in}
\end{figure}

We represent these layouts using a hierarchical data structure called a {\it Widget Tree}, where each node corresponds to a layout or interaction widget (\Cref{f:uitree}).  Layout widgets such as vertical and horizontal specify how to organize their children\footnote{Our layout widgets include: horizontal layout, vertical layout, tabs, and an adder that adds a copy of its child widget to the interface (e.g., to add multiple predicates). }, while interaction widgets\footnote{Our interaction widgets include: label, textbox, dropdown, slider, range slider, check boxes, radio buttons, and buttons.} such as Button and Dropdown are configured with subtrees (e.g., $q_1$) or values (e.g., `Sales', `Costs').

\stitle{Widgets:} We model a widget as a function $w(q, u)\to q'$, where a user interaction picks $u$ from a domain of possible values, which is then used to change the current query $q$ to a new query $q'$.  To do so, the widget replaces the subtree at a fixed path in $q$'s AST with a new subtree derived from $u$.  For example, let $q=q_1$ be the current query in \Cref{f:layouts}(a); clicking on the $q_2$ button replaces the root of $q$ with $q_2$'s AST.  In contrast, when the user selects ``Costs'' from the top drop-down in \Cref{f:layouts}(b), the \texttt{ColExpr} node in the AST will be replaced with a \texttt{ColExpr} node whose value is ``Costs''.   Similarly, clicking the toggle widget will swap between the current subtree rooted at the \texttt{WHERE} node and an empty subtree that corresponds to the absence of a \texttt{WHERE} clause in the query ($q_3$). Each widget has a fixed size only depending on the domain. For example, the button widget in \Cref{f:layouts}(c) is used to choose from the domain -- 'Sales' and 'Costs'. If the domain is larger, then there are more buttons; if it contains a longer word, the buttons will be wider. To support widgets that vary in size, we discretize the sizes and define a separate widget for each size. For example, for the button widget, we predefine small, medium and large button templates separately.

In short, each widget offers the user a choice from a domain of subtrees, and then places the chosen subtree at a widget-specific fixed location in the current AST.  The three layouts primarily differ in the paths and granularities of the subtrees that the widgets replace: layout (a) replaces the root of the current AST, whereas layouts (b) and (c) replace leaves and interior nodes of the AST. 

\vspace{-0.05in}
\subsection{The Interface Generation Problem}

Our problem is to identify changes within the input query sequence, and map them to an appropriate widget tree that can be rendered as an interactive interface.  The challenge is that the layout and the selected widgets are intertwined with the process of identifying subtree differences between the input query ASTs.  

To facilitate this process, we encode the layout and input queries in a \difftree.  Each node in the \difftree corresponds to a (possibly empty) sequence of AST nodes. There are four node types that encode differences and similarities between the input queries. \texttt{ANY} can choose one of its child nodes, \texttt{OPT} has a single child that is optional, \texttt{MULTI} has a single child that can be chosen zero or more times, and \texttt{ALL} requires all of its children to exist.  We call \texttt{ANY}, \texttt{OPT}, \texttt{MULTI} {\it choice nodes}.    Note that an AST is a special case of a \difftree, where each AST node is an \texttt{ALL} node.

A given query is expressed as the set of choices made for the choice nodes in the \difftree. For example, \Cref{f:asts} is a \difftree with the root \texttt{ANY}---choosing any of its children is equivalent to one of the input queries.    \Cref{f:diff} illustrates the \difftree for layouts \Cref{f:layouts}(b,c).  \texttt{ALL(Select)} states that all queries share the \texttt{SELECT} node as well as the \texttt{From/Table} nodes.  However, the Project clause can be chosen from \texttt{Sales} and \texttt{Costs}, and the \texttt{where} clause is optional.  Note that \Cref{f:diff} can express more queries than the initial \difftree in \Cref{f:asts}.

\begin{figure}
\centering
\includegraphics[width=.7\linewidth]{figures/difftree.pdf}
\vspace{-.1in}
\caption{\small {\difftree} for layouts \Cref{f:layouts}(b,c).}
\label{f:diff}
\vspace{-.1in}
\end{figure}

\stitle{Creating Widget Trees:} Given a \difftree, it is straightforward to derive a widget tree that can be rendered.  Each {\it choice node} is mapped to one or more interactive widgets, and \texttt{ALL} nodes are mapped to layout widgets if it contains descendant choice nodes.  

\stitle{Cost Function: }
We quantify the cost of an interface based on the usefulness and appropriateness of the widget tree $W$~\cite{Gajos2004SUPPLEAG}. Each query $q\in Q$ is expressible by selecting the appropriate values for each widget in $W$.  Thus, $U(q_i, q_{i+1}, W)$ models the minimum set of widgets that need to be changed in order to transform $q_i$ into $q_{i+1}$. $M(\cdot)$ measures whether a selected widget is well-suited for the set of subtrees it expresses.  For instance, a slider is well suited to select from a range of numeric values, but not arbitrary subtrees, whereas radio buttons are well suited for a small number of subtrees, but ill-suited for a large number.  
{\small$$C(W, Q) = \sum_{q_i\in Q} U(q_i, q_{i+1}, W) + \sum_{w\in W} M(w) $$}
For reference, \cite{Zhang2017MiningPI} only considered appropriateness when selecting widgets, and we borrow their $M(\cdot)$ cost functions.  $U(\cdot)$ accounts for the size of the minimum spanning tree that connects the widgets that need to be changed, along with the cost to interact with each of those widgets.  The cost function is a linear combination of terms that can be incrementally maintained as we explore the space of {\difftree}s and widget trees. We consider a widget tree invalid (has infinite cost) if its size exceeds the output screen's size.  
 
\section{Our Approach}

We now describe interface generation as a search problem, and our use of Monte Carlo Tree Search (MCTS)~\cite{Browne2012ASO} to efficiently search the space for a good interface.      

\vspace{-0.02in}
\subsection{Search Space}
\begin{figure}
\centering
\includegraphics[width=.85\columnwidth]{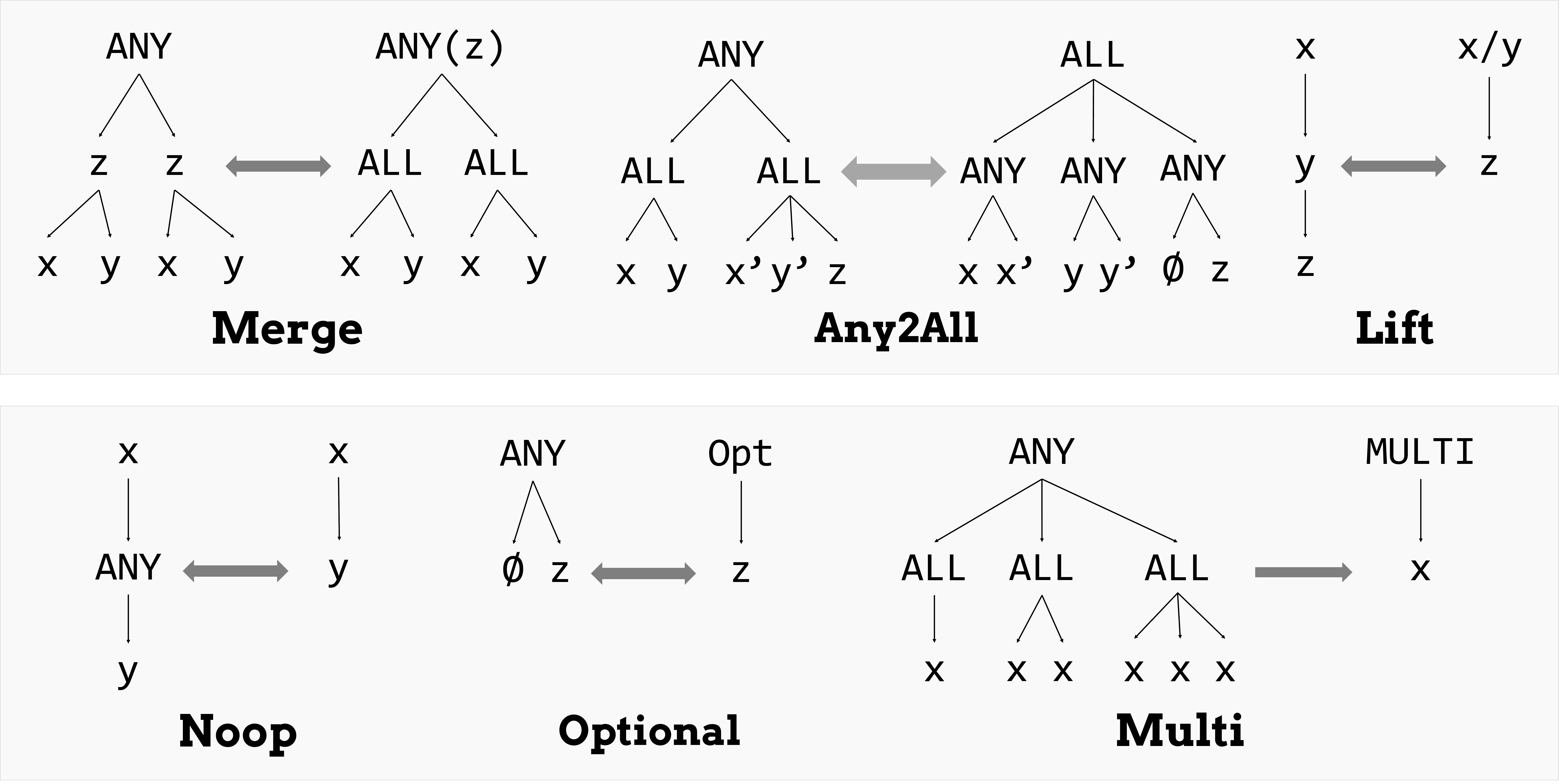}
\vspace{-.1in}

\caption{\small Set of transformation rules.}
\label{f:rule}
\vspace{-.1in}
\end{figure}

Each state in the search space is a \difftree, and the initial state is the list of input queries connected with an \texttt{ANY} node as the root.  We define a set of transition rules that transform one \difftree into another (\Cref{f:rule}). The intuition is that the initial \difftree represents a subset of the combinatorial enumeration of all expressible trees, and each rule factors out redundant substructures and variation between the trees.  In the diagram, \texttt{x,y,z} represent subtrees that are distinguished by their root nodes---the roots of \texttt{x} and \texttt{x'} are the same, and different than the root of \texttt{y}.

For instance, \texttt{Any2All} finds that \texttt{ANY} node's children all have children than can be aligned (\texttt{x}$\to$\texttt{x'}, \texttt{y}$\to$\texttt{y'}, \texttt{z}$\to\varnothing$), and groups the aligned nodes together\footnote{$\varnothing$ represents no node.}.   With the exception of \texttt{Multi}, which replaces subtrees that repeat \texttt{x} with a \texttt{MULTI} node with a single \texttt{x} child, all rules are bidirectional.  

Rules can be applied to choice nodes in the \difftree that satisfy the rule's input pattern. The number of applicable rules for the current \difftree determines the current search state's fanout. This primarily depends on the number of choice nodes and number of applicable rules for each choice node.  

For example, Listing 1 show 10 input queries. The fanout is as high as 50, and a search path can be as long as 100 steps. It is impractical to enumerate the full search space to find the lowest cost \difftree, and thus we propose the MCTS method described next. 

\begin{figure*}[bt]
\centering
\includegraphics[width=0.99\textwidth]{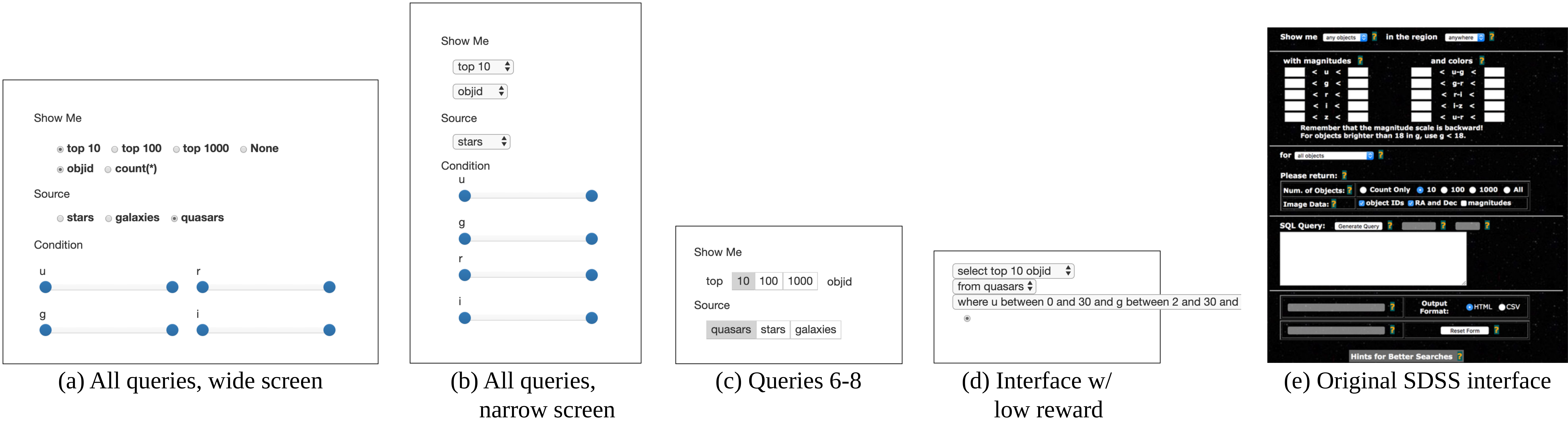}
\vspace{-.2in}
\caption{\small (a-d) Generated interfaces from queries in 
\Cref{q:qs}. (e) the pre-existing Sloan Digital Sky Survey search form.  Screenshots only show widgets and do not include the visualizations.}
\vspace{-.1in}
\label{f:exp_uiall}
\end{figure*}

\vspace{-0.02in}
\subsection{Monte Carlo Tree Search}
Monte Carlo Tree Search is used to balance exploration (trying unexplored states) with exploitation (exploring promising states) when searching in a large search space~\cite{Browne2012ASO}.  In each iteration, it performs a randomized walk of the states, and estimates the reward of the final state. It then maintains a UCT score for each visited state $s$:
{\small$$UCT^s = \frac{w^s_i}{n^s_i} + c\sqrt{\frac{\ln N^s_i}{n^s_i}}$$}
\noindent Where $w^s_i$ is the total reward for the state after the $i^{th}$ iteration.  The reward at the end of the random walk is added to every state along the path.  $n^s_i$ is the number of times the state was visited, $N^s_i$ is the number of times $s$'s parent state was visited, and $c$ is a tunable exploration parameter. 

In each iteration, we pick the state with the highest UCT, and perform a random walk of up to 200 steps from all of its immediate neighbor states. For the first iteration, we start with the initial state (e.g., Figure 1). To compute reward, we map the state (a \difftree) to the lowest cost widget tree. During the search, we randomly assign widgets to the difftree k times and select the lowest cost. The reward is the negated cost. Once the search terminates after a fixed wall clock time, we enumerate all possible widget trees for the final difftree to find the lowest cost interface.

\section{Preliminary Results}

We now present preliminary results when running our approach on the query log in \Cref{q:qs}, which is derived from the Sloan Digital Sky Survey~\cite{SDSS} query log.  We run MCTS for around 1 minute to generate each interface.  \Cref{f:exp_uiall} shows that the layout and widget selections are sensitive to the input queries and screen constraints.  (a) uses all queries in \Cref{q:qs} as input, and generates a layout for a wider screen.  It finds that the queries vary in the attributes that are selected (objid, count), as well as the number of results to return (top), and takes advantage of the wider screen to enumerate them as two sets of radio buttons.  In contrast, (b) chooses dropdown widgets due to the narrower screen.  

\Cref{f:exp_uiall}(c) shows the interface is much simpler when queries 6--8 are use as input.  These queries have the same \texttt{WHERE} clauses; since the three queries all have a \texttt{TOP} clause, the user is only asked to pick the number of rows to return (10, 100, 1000).  (d) shows a low-reward interface, and illustrates that that poor interface choices are easily possible. Finally, (e) shows the original SDSS form.

\vspace{-0.08in}
{\footnotesize\begin{lstlisting}[caption={\small Example queries used in experiments.  All queries have the same \texttt{WHERE} clause structure; for space considerations, we only show the full queries for the first two.}, label={q:qs}]
   1 select top 10 objid from stars 
     where u between 0 and 30 and g between 0 and 30 and 
           r between 0 and 30 and i between 0 and 30
   2 select top 100 objid from galaxies 
     where u between 1 and 29 and g between 10 and 30 and 
           r between 9 and 30 and i between 3 and 28 
   3 select top 1000 objid from quasars where ...
   4 select count(*) from stars where ...
   5 select objid from galaxies where ...
   6 select top 10 objid from quasars where ...
   7 select top 100 objid from stars where ...
   8 select top 1000 objid from galaxies where ...
   9 select count(*) from quasars where ...
  10 select objid from stars where ...
\end{lstlisting} 
}
\section{Ongoing Work}
 
Although we have shown that the top-down approach can generate layout-sensitive interactive interfaces, there are a number of improvements needed for it to be practically useful in terms of functionality and performance.  

A current limitation is that some combinations of widget choices may not make semantic sense;  one approach is to integrate with a query engine to benefit from its query analysis phase, another is to leverage co-occurrence of subtrees in the query log to identify likely and unlikely combinations of widget choices.  This can also inform the search phase.  Further, we are extending the widgets to support parameterized sizes---for instance, a button or dropdown can be resized depending on the available screen  space.  
 
This work has not been optimized for performance---many of the algorithms perform exhaustive enumeration, and can benefit from optimizations such as parallelization, incremental computation of the \difftree and cost functions, and search pruning.  A key optimization opportunity is to accelerate the transformation rules, which become slow to evaluate as the \difftree becomes large.  Our goal is interactive run-times.

\stitle{Acknowledgements: }  This work was supported by NSF IIS 1527765, 1564049, 1845638, and Amazon and Google awards.

{\small
\bibliographystyle{aaai}
\bibliography{main.bib}

\begin{thebibliography}{}

\bibitem[\protect\citeauthoryear{Browne \bgroup et al\mbox.\egroup
  }{2012}]{Browne2012ASO}
Browne, C.; Powley, E.~J.; Whitehouse, D.; Lucas, S.~M.; Cowling, P.~I.;
  Rohlfshagen, P.; Tavener, S.; Liebana, D.~P.; Samothrakis, S.; and Colton, S.
\newblock 2012.
\newblock A survey of monte carlo tree search methods.
\newblock {\em IEEE Transactions on Computational Intelligence and AI in Games}
  4:1--43.

\bibitem[\protect\citeauthoryear{Comber and Maltby}{1997}]{Comber1997LayoutCD}
Comber, T., and Maltby, J.~R.
\newblock 1997.
\newblock Layout complexity: Does it measure usability?
\newblock In {\em INTERACT}.

\bibitem[\protect\citeauthoryear{Gajos and Weld}{2004}]{Gajos2004SUPPLEAG}
Gajos, K.~Z., and Weld, D.~S.
\newblock 2004.
\newblock Supple: automatically generating user interfaces.
\newblock In {\em IUI}.

\bibitem[\protect\citeauthoryear{Mackinlay, Hanrahan, and
  Stolte}{2007}]{Mackinlay2007ShowMA}
Mackinlay, J.~D.; Hanrahan, P.; and Stolte, C.
\newblock 2007.
\newblock Show me: Automatic presentation for visual analysis.
\newblock {\em IEEE Transactions on Visualization and Computer Graphics} 13.

\bibitem[\protect\citeauthoryear{SDSS}{2017}]{SDSS}
SDSS.
\newblock 2017.
\newblock Sloan digital sky survey.
\newblock http://www.sdss.org/.

\bibitem[\protect\citeauthoryear{Sievert \bgroup et al\mbox.\egroup
  }{2017}]{sievert2017plotly}
Sievert, C.; Parmer, C.; Hocking, T.; Chamberlain, S.; Ram, K.; Corvellec, M.;
  and Despouy, P.
\newblock 2017.
\newblock plotly: Create interactive web graphics via ‘plotly. js’.
\newblock {\em R package version} 4(1):110.

\bibitem[\protect\citeauthoryear{Zhang, Sellam, and
  Wu}{2017}]{Zhang2017MiningPI}
Zhang, H.; Sellam, T.; and Wu, E.
\newblock 2017.
\newblock Mining precision interfaces from query logs.
\newblock In {\em SIGMOD Conference}.

\end{thebibliography}
\appendix
}

\end{document}